\documentclass[cameraready]{Interspeech}

\title{LibriTTS-VI: A Public Corpus and Novel Methods \\for Efficient Voice Impression Control}

\author[affiliation={1}, orcid=0009-0002-0342-0214]{Junki}{Ohmura}
\author[affiliation={1}, orcid=0009-0001-5565-7240]{Yuki}{Ito}
\author[affiliation={1}, orcid=0009-0003-1776-9668]{Emiru}{Tsunoo}
\author[affiliation={1}, orcid=0009-0004-1407-7401]{Toshiyuki}{Sekiya}
\author[affiliation={1}, orcid=0009-0006-1939-4988]{Toshiyuki}{Kumakura}

\address{
    $^1$ Sony Group Corporation, Japan
}

\email{\{junki.ohmura, yuki.g.ito, emiru.tsunoo, toshiyuki.sekiya, toshiyuki.kumakura\}@sony.com}


\usepackage{comment}
\usepackage{multirow}
\usepackage{makecell}
\usepackage{float}
\usepackage{caption}

\begin{document}

\maketitle

\begin{abstract}
Numerical voice impression (VI) control (e.g., scaling brightness) enables fine-grained control in text-to-speech (TTS).
However, it faces two challenges: no public corpus and impression leakage, where reference audio biases synthesized voice away from the target VI.
To address the first challenge, we introduce LibriTTS-VI, the first public VI corpus built on LibriTTS-R.
For the second, we hypothesize a single reference causes leakage by entangling speaker identity and VI.
To mitigate this, we propose 1) disentangled training with two utterances from the same speaker for speaker and VI conditioning, and 2) a reference-free method controlling the impression solely via target VI.
Experimentally, our best method improves controllability: 11-dimensional VI mean squared error drops from 0.61 to 0.41 objectively and 1.15 to 0.92 subjectively.
A comparison with a prompt-based TTS reveals imprecise numerical control and entanglement between VI and text semantics, which our methods overcome.\footnote{The corpus and demo samples are publicly available. \url{https://github.com/sony/LibriTTS-VI}}
\end{abstract}

\setlength{\abovedisplayskip}{2pt} \setlength{\belowdisplayskip}{2pt}

\vspace{-0.2cm}
\section{Introduction}
\label{sec:intro}
\vspace{-0.15cm}

Modern text-to-speech (TTS) synthesis has achieved near-human naturalness \cite{taco2, fs2, vits}, shifting the research toward controlling speaker identity and speaking styles.
Control paradigms include manipulating acoustic features such as pitch, energy, and duration \cite{fs2, vits, drawspeech}, and ID-based methods for predefined speakers \cite{vits, kong23_vits2_interspeech} or emotional classes \cite{lee2017emotional, diatlova23_emospeech_ssw}. 
Another prominent paradigm involves conditioning on audio exemplars or text descriptions, utilizing reference-based speaker encoders \cite{jia2018transfer, pmlr-yourtts}, unsupervised style transfer \cite{wang2018style}, emotion synthesis with intensity control \cite{cho24_emosphere, cho25_emosphere++}, and the use of natural-language (NL) prompts for prosody, emotion, or speaker traits \cite{guo2023prompttts,liu2023promptstyle,yang2024instructtts,shimizu2024prompttts++,kawamura2024librittsp,sigurgeirsson2024controllable,jeon2025prompt}.
Recent large language model (LLM)-based models have further advanced this direction, enabling complex NL instruction \cite{ zhou2024voxinstruct, hu2026voicesculptor, Qwen3-TTS}.
Yet controllability often trades off precision and usability.
While manipulating acoustic features offers precise control, it is too granular and difficult for users.
Other intuitive approaches introduce different constraints: IDs and references are limited to predefined categories or available exemplars, while NL prompts lack fine-grained numerical control.

A key challenge is balancing fine-grained control with human-understandable scales for practical use.
One line of work derives controllable axes from speaker-embedding spaces \cite{vanrijn22_interspeech}, which was extended in \cite{lux23_interspeech} to generate reference-free artificial speaker embeddings.
Another paradigm uses explicitly defined perceptual dimensions for voice characteristics, categorized into two approaches.
One approach uses perceptual voice qualities (e.g., roughness and breathiness), rated by phonetic experts based on clinical protocols \cite{rautenberg2025speech}.
Another approach, voice impression control (VIC) \cite{fujita2025vic}, targets voice impressions (VI) such as calmness and brightness based on non-experts' subjective ratings \cite{kido1999extraction, nagano2025influence}, which we focus on in this paper. 

VIC \cite{fujita2025vic} enables VI-based numerical control of speaker characteristics in TTS with 11 VI dimensions.
However, VIC still faces two limitations.
First, it relies on a private VI corpus, making follow-up studies difficult.
Second, we observed impression leakage: despite being able to specify the reference audio and target VI individually, the synthesized voice is biased toward the impression of the reference audio itself.

This paper addresses these two challenges.
We address the first challenge by introducing LibriTTS-VI, a public VI corpus built on LibriTTS-R \cite{koizumi23_interspeech}.
For the second challenge, we hypothesize that training with one utterance for both speaker and VI conditioning causes impression leakage.
Therefore, we propose two methods to mitigate the leakage.
The first one is a disentangled training strategy using two utterances from the same speaker to individually provide speaker identity and the target VI.
The second one is a reference-free method controlling the impression only from the target VI without reference audio.
Our best method objectively reduced the mean squared error (MSE) of 11-dimensional VI vectors from 0.61 to 0.41 and from 1.15 to 0.92 subjectively, while maintaining high synthesis quality.
We also evaluate a recent LLM-based TTS, revealing its imprecise numerical control and entanglement between VI and text semantics, which our methods effectively address.

\vspace{-0.2cm}
\section{VIC Task Definition}
\label{sec:vic_overview}
\vspace{-0.15cm}

\begin{figure*}[t!]
    \centering
    \includegraphics[width=0.891\textwidth]{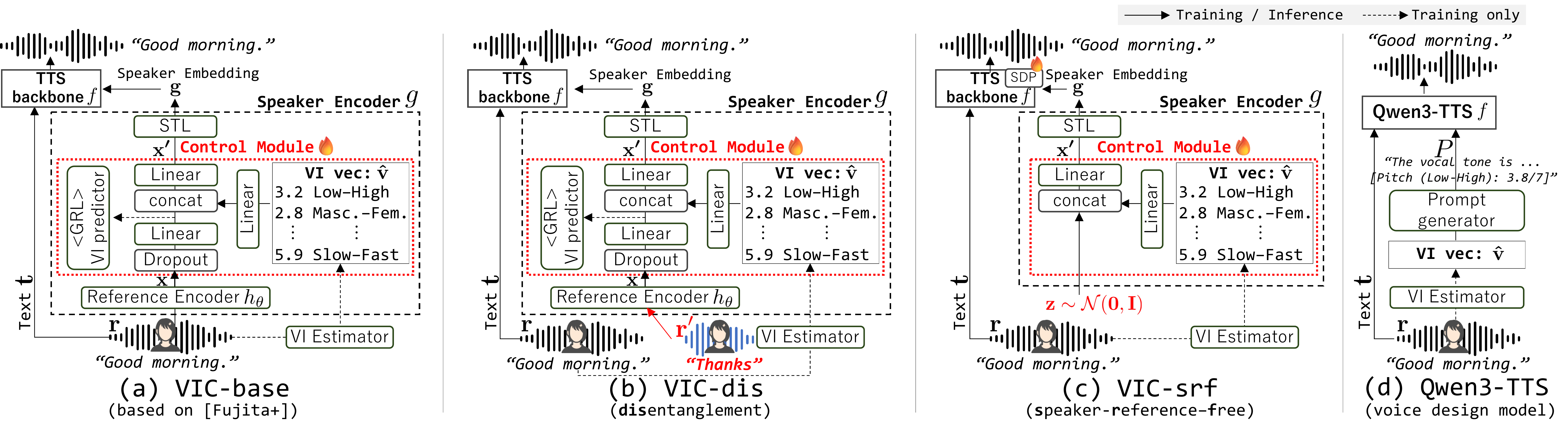}
    \vspace{-0.25cm}
    \caption{Overview of the VIC systems (base, dis, and srf) and the Qwen3-TTS voice design interface for VIC.}
    \label{fig:methods}
    \vspace{-0.53cm}
\end{figure*}

VIC \cite{fujita2025vic} synthesizes speech that follows a target VI while preserving the speaker traits of a reference utterance.
The target VI is an 11-dimensional vector $\mathbf{v} \in \mathbb{R}^{11}$ on a seven-point scale (1--7): A) Low--High pitched, B) Masculine--Feminine, C) Clear--Hoarse, D) Calm--Restless, E) Powerful--Weak, F) Youthful--Aged, G) Thick--Thin, H) Firm--Relaxed, I) Dark--Bright, J) Cold--Warm, and K) Slow--Fast.\footnote{For a fair comparison, we retained sensitive scales (e.g., Masculine-Feminine) with minor wording and polarity adjustments for clarity.}
Formally, VIC outputs waveform $\mathbf{y}$ from text $\mathbf{t}$, reference utterance $\mathbf{r}$, and target VI $\mathbf{v}$:
\begin{equation}
    \mathbf{y} = f(\mathbf{t}, \mathbf{g}) = f(\mathbf{t}, g(\mathbf{r}, \mathbf{v})),
    \label{eq:vic}
\end{equation}
where $f$ is a TTS backbone and $g$ is a speaker encoder extracting the speaker embedding $\mathbf{g}$ aligned with $\mathbf{v}$.
The speaker encoder $g$ (black dashed box in Fig.~\ref{fig:methods} (a)) comprises a reference encoder $h_\theta$ with parameters $\theta$, a control module (CM) for VI, and a style token layer (STL) \cite{wang2018style}.
The reference encoder consists of a pre-trained HuBERT \cite{hsu2021hubert} and a Bi-LSTM with attention.
It encodes a reference $\mathbf{r}$ into an intermediate vector $\mathbf{x}$:
\begin{equation}
    \mathbf{x} = h_\theta(\mathbf{r}) = \text{BiLSTM}_{\text{attn}}(\text{HuBERT}(\mathbf{r})).
    \label{eq:intermediate_vec}
\end{equation}
CM modulates $\mathbf{x}$ into $\mathbf{x'}$ given the target VI $\mathbf{v}$.
Specifically, it processes $\mathbf{x}$ with a high-rate dropout and a linear layer, and $\mathbf{v}$ via a linear layer.
These outputs are concatenated and linearly projected to match the dimensionality of $\mathbf{x}$, yielding $\mathbf{x'}$:
\begin{equation}
    \mathbf{x'} = \text{Linear}([\text{Linear}(\text{Dropout}(\mathbf{x})); \text{Linear}(\mathbf{v})]),
    \label{eq:x_prime}
\end{equation}
where $[\cdot; \cdot]$ is a concatenation operation. 
This vector is then passed through the STL to produce the speaker embedding $\mathbf{g}$:
\begin{equation}
    \mathbf{g} = g(\mathbf{r}, \mathbf{v}) = \text{STL}(\mathbf{x'}).
    \label{eq:speaker_emb}
\end{equation}
Note that the CM is inserted into $g$ only after jointly pre-training the backbone $f$ and the speaker encoder $g$ without the CM.
The CM is trained with all other modules frozen, using the same training objectives of the backbone TTS to implicitly learn the correspondence between a ground-truth (GT) audio and its VI.

During the training, a pre-trained VI estimator (VIE) predicts target $\mathbf{\hat{v}}$ from reference $\mathbf{r}$.
VIE shares the architecture of $h_\theta$ (with independent parameters $\theta'$) followed by a linear head:
\begin{equation}
\mathbf{\hat{v}} = \text{VIE}(\mathbf{r}) = \text{Linear}(h_{\theta'}(\mathbf{r})).
\label{eq:vi-estimator}
\end{equation}
By setting $\mathbf{v} = \mathbf{\hat{v}}$, Eq.~(\ref{eq:vic}) becomes:
\begin{equation}
 \mathbf{y} = f(\mathbf{t}, g(\mathbf{r}, \mathbf{\hat{v}})) = f(\mathbf{t}, g(\mathbf{r}, \text{VIE}(\mathbf{r}))).
\label{eq:vic-base}
\end{equation}
Because both $\mathbf{x}$ and $\mathbf{\hat{v}}$ come from the same reference $\mathbf{r}$, speaker identity and VI tend to be entangled.
To encourage disentanglement, a VI predictor consisting of two linear layers with a gradient reversal layer (GRL) \cite{ganin2015unsupervised} is applied to the first term in the concatenation of Eq.~(\ref{eq:x_prime}).
This suppresses VI cues in $\mathbf{x}$, mitigating impression leakage.

Despite the success of VIC \cite{fujita2025vic}, two challenges remain: the lack of a public VI corpus and impression leakage.
To address the first challenge, we introduce our new VI corpus, LibriTTS-VI, in Sec.~\ref{sec:dataset}.
The second challenge is impression leakage, which we still observed in our preliminary experiments despite employing the high-rate dropout and the GRL.
We hypothesize that impression leakage arises because a single reference utterance, $\mathbf{r}$, is used for both speaker and VI conditioning during training.
To this end, we propose two methods in Sec.~\ref{sec:proposed_methods}.

\vspace{-0.2cm}
\section{LibriTTS-VI: A Public Corpus for VIC}
\label{sec:dataset}
\vspace{-0.15cm}

To foster reproducibility, we created LibriTTS-VI, a new public corpus, built by annotating LibriTTS-R \cite{koizumi23_interspeech}.
The released dataset contains manual VI annotations, the annotation guideline, and estimated VI values for the entire LibriTTS-R corpus.

\begin{figure}[t]
    \centering
    \includegraphics[width=\linewidth]{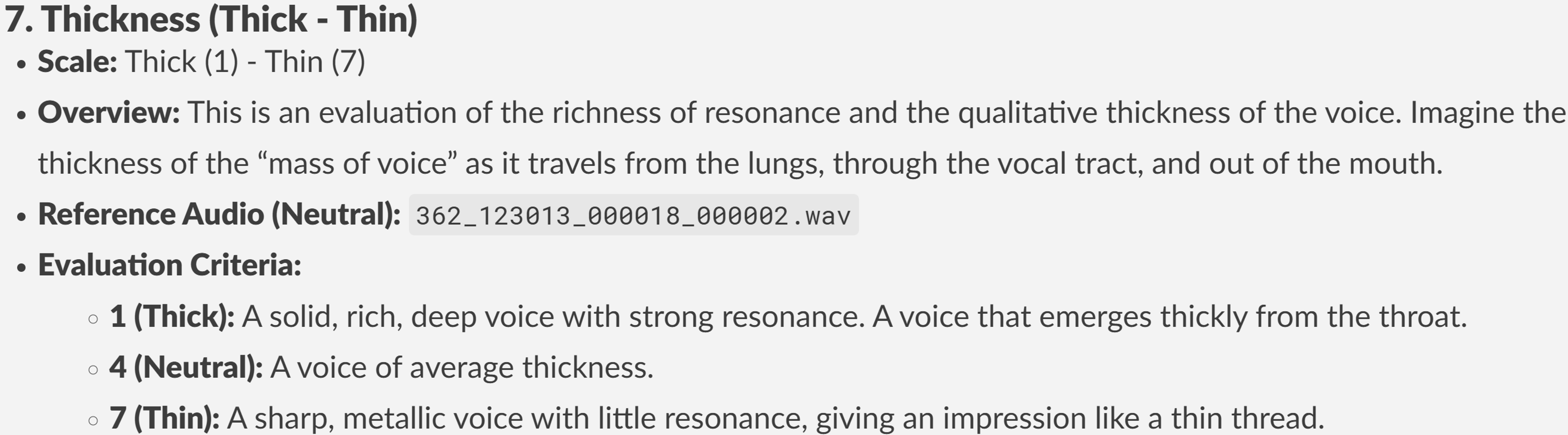}
    \vspace{-0.6cm}
    \caption{An example of VI annotation guideline.}
    \label{fig:anno}
    \vspace{0.1cm}
\captionof{table}{Krippendorff's alpha for the annotated VI dimensions.}
\vspace{-0.35cm}
\label{tab:krippendorff_alpha_multi}
\footnotesize
\renewcommand{\arraystretch}{0.9}
\begin{tabular}{@{}l@{\hspace{1.1mm}}c@{\hspace{2.7mm}}l@{\hspace{1.1mm}}c@{\hspace{2.7mm}}l@{\hspace{1.1mm}}c@{}}
\toprule
\textbf{VI Dim.} & \textbf{$\alpha$} & \textbf{VI Dim.} & \textbf{$\alpha$} & \textbf{VI Dim.} & \textbf{$\alpha$} \\
\midrule
A) Low--High     & .729 & E) Pow.--Weak    & .331 & I) Dark--Bright & .433 \\
B) Masc.--Fem.   & .888 & F) Youth.--Aged  & .586 & J) Cold--Warm   & .197 \\
C) Clear--Hoarse & .415 & G) Thick--Thin   & .413 & K) Slow--Fast   & -    \\
D) Calm--Rest.   & .295 & H) Firm--Relax.  & .408 & \textbf{Average}    & .470 \\
\bottomrule
\end{tabular}
\vspace{-0.6cm}
\end{figure}

For annotation, we randomly selected 130 utterances from distinct speakers in the LibriTTS-R training set.
We defined criteria for the ten VIs other than K) Slow--Fast, using written descriptions of each VI and reference audio as an anchor sample, as exemplified in Fig.~\ref{fig:anno}.
Four in-house experienced annotators rated the 130 utterances on a seven-point Likert scale.
We used the mean of the four ratings as the final VI score, yielding 1,300 labels (130 $\times$ 10 VIs).\footnote{We are currently expanding the annotations to 300 utterances.}
For K) Slow--Fast, we computed words-per-minute via timestamped automatic speech recognition (ASR) \cite{lintoai2023whispertimestamped}, and rescaled to 1--7.
Table~\ref{tab:krippendorff_alpha_multi} reports inter-annotator agreement by Krippendorff's alpha.
Although many scales fall below the reliable threshold ($\alpha \ge 0.667$) \cite{krippendorff2018}, our average (0.470) exceeds similar vocal perception tasks: speech emotion (0.442) \cite{perepelkina2018ramas} and singing voices (0.153) \cite{bruder2024perceptual}.
Moreover, \cite{bruder2024perceptual} showed such low-agreement ratings remain good predictors of voice preference, supporting the value of our dataset.

As in \cite{fujita2025vic}, we trained VIE in Eq.~(\ref{eq:vi-estimator}) to label the rest of the LibriTTS-R corpus to scale the annotation.
The VIE is trained with MSE against manual VI labels.
With limited 130 labeled utterances, we augmented data to train the VIE, as in \cite{fujita2025vic}.
The VIE is trained by assigning the same manual label to all utterances of a speaker, assuming a constant VI per speaker.
However, this assumption is too strict for LibriTTS-R, which is a mixture of narration and expressive utterances.
Therefore, we adopted a new data augmentation strategy.
To safely assign the manual labels to other utterances of the same speaker, we select acoustically similar utterances, assuming similar VI.
The similarity was determined by averaging ranks of L1 distances of pitch and energy and cosine similarities of the WavLM embeddings \cite{chen2022wavlm, barrault2023seamless}.
For each manual label, we only used the 100 most similar utterances from the same speaker, resulting a 100-fold data augmentation.
Of the 130 manually annotated utterances (1,300 VI labels), we used 100 for augmentation to train the VIE, and held out 30 for the test set. 
The VIE achieved a root-MSE of 0.68 ($<1$ point) on this set.

\vspace{-0.3cm}
\section{Proposed Methods}
\label{sec:proposed_methods}
\vspace{-0.15cm}
We hypothesize impression leakage arises because the same reference $\mathbf{r}$ conditions both speaker and VI, as in Eq.~(\ref{eq:vic-base}).
As a result, the speaker encoder $g$ encodes VI cues from the reference $\mathbf{r}$ alongside speaker traits.
This can be prevented by disentangling VI and speaker information extracted from the reference utterance.
Furthermore, if the VI sufficiently represents speaker identity, we can omit $\mathbf{r}$ and rely solely on $\mathbf{v}$.

\subsection{Disentanglement via different utterances (VIC-dis)}
\label{ssec:vic_dis}
\vspace{-0.15cm}

VIC-dis (``disentanglement''), shown in Fig.~\ref{fig:methods} (b), tackles this entanglement through a decoupled training strategy.
Since extracting speaker identity does not require the reference audio to match the target VI, we can source these inputs individually.
During training, we introduce another utterance $\mathbf{r'}$ from the same speaker, to replace $\mathbf{r}$ in Eq.~(\ref{eq:vic}).
The original utterance $\mathbf{r}$ remains as the GT synthesis target and the target VI is also extracted from the same $\mathbf{r}$.
This inherent VI difference between $\mathbf{r'}$ and $\mathbf{r}$ decouples identity and VI without architectural changes.
The training of VIC-dis reformulates Eq.~(\ref{eq:vic-base}) as:
\vspace{-0.05cm}
\begin{equation}
\label{eq:vic-dis}
\mathbf{y} = f(\mathbf{t}, g(\mathbf{r'}, \text{VIE}(\mathbf{r}))).
\end{equation}

\vspace{-0.35cm}
\subsection{Speaker-reference-free generation (VIC-srf)}
\label{ssec:vic_srf}
\vspace{-0.15cm}
VIC-srf (``speaker-reference-free''), shown in Fig.~\ref{fig:methods} (c), architecturally eliminates the impression leakage by removing the speaker reference from the synthesis process.
Inspired by the method of generating pseudo speaker embeddings \cite{lux23_interspeech}, the synthesis is conditioned solely on the target VI vector by replacing the first concatenation term in Eq.~(\ref{eq:x_prime}) with Gaussian noise $\mathbf{z} \sim \mathcal{N}(\mathbf{0}, \mathbf{I})$.
The training of VIC-srf reformulates Eq.~(\ref{eq:vic-base}) as:
\vspace{-0.05cm}
\begin{equation}
\label{eq:vic-srf}
    \mathbf{y} = f(\mathbf{t}, g(\mathbf{z}, \text{VIE}(\mathbf{r}))).
\end{equation}

\vspace{-0.35cm}
\section{Experiments}
\label{sec:experiments}
\vspace{-0.15cm}

We objectively and subjectively evaluated controllability and synthesis quality on the LibriTTS-R test-clean set, enabling zero-shot evaluation on 39 unseen speakers.
We evaluated five VIC models: the original VIC (VIC-base) \cite{fujita2025vic}, our proposed VIC-dis and VIC-srf, and two LLM-based VIC variants.

\vspace{-0.3cm}
\subsection{Experimental setup}
\label{ssec:experimental_setup}
\vspace{-0.2cm}

To improve audio quality, we replaced the FastSpeech2 \cite{fs2} backbone ($f$ in Eq.~(\ref{eq:vic})) in \cite{fujita2025vic} with VITS \cite{vits}. 
Our baseline, VIC-base (Fig.~\ref{fig:methods} (a)), uses VITS with a single reference $\mathbf{r}$ for speaker and VI conditioning.
To test if NL prompts can achieve VIC, we evaluated the state-of-the-art Qwen3-TTS voice design model\footnote{\texttt{Qwen3-TTS-12Hz-1.7B-VoiceDesign}} (QVD) \cite{Qwen3-TTS, huang2025instructttseval} (Fig.~\ref{fig:methods} (d)).
Because QVD uses only NL prompts for speaker conditioning, we built a prompt generator mapping a VI to an NL prompt $P$, inspired by LLM-based vocal attribute-to-text methods \cite{parler_tts, huang2025instructttseval}.
To map the continuous VI value $v_d$ of each dimension $d \in \{1,\dots,11\}$ to text, we quantized it into a discrete level $l \in \{1,\dots,7\}$.
For each $(d, l)$, we prompted Gemini 3 Pro \cite{googlegemini3} with our annotation guidelines to generate five paraphrases $\mathcal{D}_{d,l}=\{s^{(1)}_{d,l},\dots,s^{(5)}_{d,l}\}$. 
To mitigate quantization error ($v_d$ to $l$), we appended $v_d$ to each sampled sentence as a tag $\phi(v_d)$ (e.g., ``\textit{[Pitch (Low–High): $3.2/7$]}'').
Given a VI $\mathbf{v}$, we sampled one sentence per dimension and concatenated them to form the prompt $P$:
\begin{equation}
P = \bigoplus_{d=1}^{11} \bigl( s_d \oplus \phi(v_d) \bigr),
\quad
s_d \sim \mathrm{Uniform}(\mathcal{D}_{d,\mathrm{round}(v_d)}),
\label{eq:prompt_gen}
\end{equation}
where $\oplus$ denotes string concatenation with a space delimiter.
We evaluated two settings: \textbf{QVD-z (zero-shot)} used the pre-trained model to test zero-shot capability for VIC, and \textbf{QVD-f (fine-tuned)} was fine-tuned on LibriTTS-R with prompts $P$.

Our VITS backbone was based on a public implementation.\footnote{\url{https://github.com/daniilrobnikov/vits2}}
To further improve synthesis quality, we added a Conformer-based text encoder \cite{gulati20_interspeech,hayashi2021espnet2}, and applied a connectionist temporal classification loss and an adaptive layer-norm zero Transformer, following  \cite{hierspeech,hierspeech++,peebles2023scalable}.
We set the dimension of $\mathbf{x}$ and $\mathbf{g}$ to 256.
The VITS backbone was pre-trained for 600k steps, followed by 60k steps of CM fine-tuning where $\mathbf{v}$ and $\mathbf{x}$ were projected to 32 dimensions.
For VIC-srf, we fine-tuned the CM and the stochastic duration predictor (SDP) of VITS to stabilize the speaking rate.
QVD-f was fine-tuned with $P$ as voice design instructions for 5 epochs.
All models (except QVD-z) were trained on eight NVIDIA A100 GPUs using AdamW.
We used a batch size of 20 and lr of $2\times10^{-4}$ for VITS-based systems, and a batch size of 2  (gradient accumulation 4) and lr of $1\times10^{-6}$ for QVD-f.

\vspace{-0.25cm}
\subsection{Objective evaluation}
\label{ssec:obj_results}
\vspace{-0.2cm}

\begin{table}[t]
\centering
\footnotesize
\renewcommand{\arraystretch}{0.95}
\setlength{\tabcolsep}{0.4pt} 
\caption{Summary of objective evaluation results. We measured intelligibility with CER and WER, audio quality with UTMOS, speaker similarity with SECS, and VI control error with VI-MSE, RVI-MSE, and impression leakage with their difference ($\Delta_{\text{V}}$). $^\dagger$ indicates a statistically significant improvement of VIC-dis or VIC-srf over VIC-base ($p < 0.05$). * indicates that $\Delta_{\text{V}}$ is significantly different from zero ($p<0.05$).}
\label{tab:exp_comparison}
\vspace{-0.3cm}
\begin{tabular}{@{}l|c|c|c|c|c|c|c@{}} 
\hline
\textbf{Model} & \textbf{CER} & \textbf{WER} & \textbf{UTMOS} & \textbf{SECS} & \textbf{VI-MSE} & \textbf{RVI-MSE} & \textbf{$\Delta_{\text{V}}$} \\
\hline 
\textbf{GT} & 2.41 & 6.17 & 4.17 & \makecell{0.81 (same)\\0.63 (cross)} & - & - & - \\
\hline 
\textbf{VITS} & 3.36 & 8.63 & 4.20 & \textbf{0.77} & - & - & - \\
\textbf{VIC-base} & 3.20 & 8.17 & 4.23 & \textbf{0.77} & 0.39 & 0.61 & 0.22* \\
\textbf{VIC-dis} & 3.12$^\dagger$ & 7.99 & 4.25$^\dagger$ & 0.75 & 0.37 & 0.51$^\dagger$ & 0.14* \\
\textbf{VIC-srf} & \textbf{2.99}$^\dagger$ & \textbf{7.72}$^\dagger$ & \textbf{4.26}$^\dagger$ & 0.72 & \textbf{0.36}$^\dagger$ & \textbf{0.41}$^\dagger$  & \textbf{0.05}  \\
\hline 
\textbf{QVD-z} & \textbf{1.95} & \textbf{5.42} & 4.27 & 0.58 & \textbf{0.82} & \textbf{0.97} & \textbf{0.15}* \\
\textbf{QVD-f}  & 2.19 & 6.02 & \textbf{4.37} & \textbf{0.65} & 0.87 & 1.19 & 0.32* \\
\hline
\end{tabular}
\vspace{-0.72cm}
\end{table}

Objective evaluation used all 4,837 utterances from the test-clean set.
We measured speech intelligibility with character/word error rates (CER/WER) via Whisper large-v3 \cite{radford2023robust}, audio quality with UTMOS \cite{saeki22c_interspeech}, and speaker similarity with speaker encoder cosine similarity (SECS) \cite{louppe2019resemblyzer}.
For fair SECS comparison, we compared each sample to all the utterances from the same speaker but the GT used as the reference $\mathbf{r}$.
As reference bounds, we report SECS (same/cross) from all same- and cross-speaker original speech pairs.
To assess controllability, we defined two metrices; VI-MSE and RVI-MSE.
VI-MSE is the MSE between the predicted VI from the synthesized speech and the target VI $\mathbf{v_r}$ from the speaker reference $\mathbf{r}$, while RVI-MSE uses a target $\mathbf{v_*}$ extracted from a randomly sampled utterance of a different speaker:
\begin{equation}
    \text{VI-MSE} = || \mathbf{v_r} - \text{VIE}(f(\mathbf{t}, g(\mathbf{r}, \mathbf{v_r}))) ||_2^2,
    \label{eq:vi_mse}
\end{equation}
\begin{equation}
    \text{RVI-MSE} = || \mathbf{v}_{*} - \text{VIE}(f(\mathbf{t}, g(\mathbf{r}, \mathbf{v}_{*}))) ||_2^2,
    \label{eq:rvi_mse}
\end{equation}
where the speaker reference $\mathbf{r}$ is omitted for VIC-srf and QVD.
We quantify this by $\Delta_{\text{V}} = \text{RVI-MSE} - \text{VI-MSE}$.
If there is no VI leakage from the reference $\mathbf{r}$, $\Delta_{\text{V}}$ should be zero. 

Table \ref{tab:exp_comparison} shows the results.
The VITS-based systems (VIC-base/dis/srf) maintained backbone performance; compared to GT, they showed marginally better UTMOS but slightly worse WERs and CERs.
SECS stayed high for reference-based models (VIC-base/dis), and VIC-srf remained above SECS (cross) (0.72 vs. 0.63).
Our proposed methods significantly reduced leakage ($\Delta_{\text{V}}$) from 0.22 (VIC-base) to 0.14 (dis) and 0.05 (srf), with VIC-srf showing no significant VI-MSE/RVI-MSE difference.
This confirmed its structural leakage elimination.
For QVD models, despite strong ASR/UTMOS, they struggled with speaker identification and VI control; their SECS fell below the SECS (cross) (0.58 vs. 0.63), with marginal recovery even after fine-tuning (0.65).
High VI-MSE indicates NL prompts lack fine-grained precision, and QVD-f further worsened it.
This suggests that adapting large pre-trained models to VIC is non-trivial.
QVD models also exhibited large $\Delta_{\text{V}}$ despite being speaker-reference-free.
We found QVD's output VI was sensitive to text content.
For instance, including an exclamation mark in the text biased the predicted VI toward restlessness.
This suggests entanglement of content and VI, where text semantics bias prosody in LLM-TTS \cite{lajszczak2024base}.

To quantify this, we trained a Ridge regressor per VI dimension to predict VI values from text embeddings \cite{reimers-2019-sentence-bert}.\footnote{We used \texttt{all-MiniLM-L6-v2} model.}
We then calculated the correlation between these text-predicted VIs and the audio-estimated VIs from VIE (Eq.~(\ref{eq:vi-estimator})) as an entanglement proxy.
We found that correlations exceeded $0.2$ for five dimensions, suggesting text semantics align with certain VI attributes.
In the RVI-MSE setting where the target VI can be misaligned with the text semantics, QVD struggled to decouple them; introducing a semantic bias into the output VI.

\begin{table}[t]
\renewcommand{\arraystretch}{0.6}
\caption{Slopes of linear fits for target vs. predicted VI values in the modulation experiment. Larger positive slopes indicate higher responsiveness. Bold: best, underline: second-best.}
\label{tab:slopes}
\footnotesize
\vspace{-0.3cm}
\setlength{\tabcolsep}{2.5pt} 
\begin{tabular}{@{}lcccc|lcccc@{}}
\toprule
\textbf{VI} & \textbf{base} & \textbf{dis} & \textbf{srf} & \textbf{QVD-z} & \textbf{VI} & \textbf{base} & \textbf{dis} & \textbf{srf} & \textbf{QVD-z} \\
\midrule
A) L-H & .008 & \underline{.106} & \textbf{.145} & .035 & G) T-T & .037 & \underline{.043} & \textbf{.060} & -.008 \\
B) M-F & \textbf{.858} & \underline{.856} & .843 & .534 & H) F-R & .062 & \underline{.090} & \textbf{.142} & .026 \\
C) C-H & .042 & \textbf{.062} & \underline{.050} & .028 & I) D-B & .035 & \underline{.044} & \textbf{.120} & .023 \\
D) C-R & .064 & \underline{.149} & \textbf{.179} & .009 & J) C-W & .045 & \underline{.102} & \textbf{.147} & .000 \\
E) P-W & -.037 & -.013 & \textbf{.014} & \underline{.008} & K) S-F & .011 & \underline{.043} & \textbf{.205} & .011 \\
F) Y-A & .207 & \underline{.273} & \textbf{.280} & .075 & \textbf{Average} & .121 & \underline{.159} & \textbf{.199} & .068 \\
\bottomrule

\end{tabular}
\vspace{-0.45cm}
\end{table}

\begin{figure}[t]
    \centering
    \includegraphics[width=0.8\linewidth]{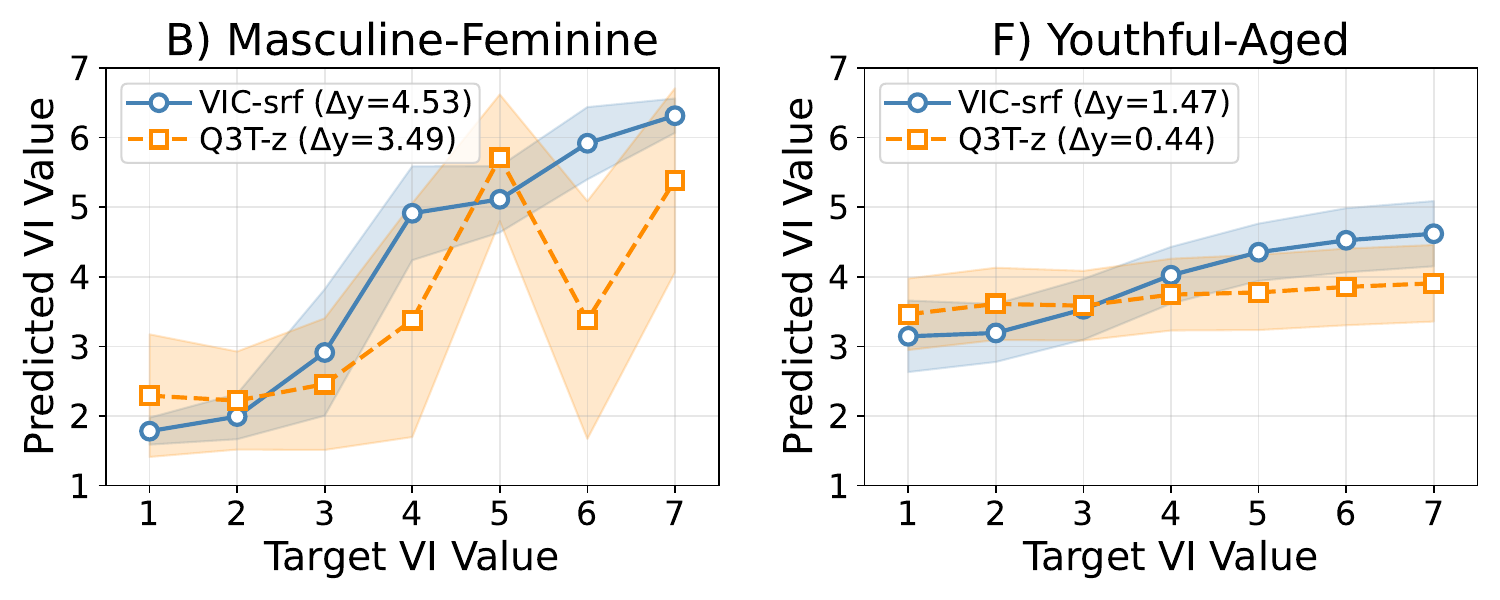}
    \vspace{-0.43cm}
    \caption{VI modulation experiment for B) Masculine-Feminine and F) Youthful-Aged dimensions. Shaded areas: $\pm 1\sigma$.}
    \label{fig:slope_comparison}
    \vspace{-0.76cm}
\end{figure}

To further probe control precision, we ran a VI modulation experiment following \cite{fujita2025vic}.
For all the test speakers, we modulated single VI dimensions from 1 to 7 in steps of 1 while keeping the others fixed.
For each condition, we synthesized 10 sentences and predicted VI with VIE in Eq.~(\ref{eq:vi-estimator}).
We then quantified the control fidelity via the slope of the linear fit between the target and predicted VI.
Larger positive slopes indicate more responsive control in the intended direction.
Table \ref{tab:slopes} shows a consistent hierarchy for VITS-based systems: VIC-srf (Avg: 0.199) $>$ dis (0.159) $>$ base (0.121).
On E) Powerful--Weak, VIC-base/dis yielded negative slopes, whereas VIC-srf remains positive (0.008).
Because QVD-f showed lower VI controllability, we evaluated only QVD-z.
Even in zero-shot, QVD-z yields positive slopes for all except G) Thick--Thin, but is less sensitive (Avg: 0.068) than VITS-based systems.
Fig.~\ref{fig:slope_comparison} shows QVD-z's instability: it fluctuates on B) Masculine--Feminine, while F) Youthful--Aged yields a narrower prediction range than VIC-srf ($0.44$ vs. $1.47$ points).
This imprecise numerical control via NL descriptions aligns with prior findings in LLMs \cite{li-etal-2025-exposing}.

\vspace{-0.2cm}
\subsection{Subjective evaluation}
\label{ssec:subj_results}
\vspace{-0.1cm}

We ran subjective tests of controllability and audio quality on only VITS-based systems.
We evaluated single-VI and multiple-VI modulation settings.
For the single-VI modulation, following \cite{fujita2025vic}, we shifted four representative VIs by $\pm 3$: A) Low--High\footnote{We replaced the Powerful--Weak dimension from the original VIC study with Low-High, as our models struggled to learn the former.}, B) Masculine--Feminine, F) Youthful--Aged, and I) Dark--Bright.
For the multiple-VI modulation, we assessed the simultaneous modulation of multiple VIs.
We reused the same synthesized utterances from the RVI-MSE evaluation (Sec.~\ref{ssec:obj_results}, Table~\ref{tab:exp_comparison}).
To cover diverse VI modulation settings at practical cost, we evaluated one male and one female speaker (IDs 1089 and 8555).
We only report ID~8555 as a representative case because the main trends were consistent across the two speakers.

First, we evaluated controllability.
For each system, we selected 10 sentences for each single-VI shift ($\pm3$) and for the multiple-VI (RVI-MSE) setting.
The same four annotators as in Sec.~\ref{sec:dataset} rated the VI of the synthesized speech, with each annotator completing 1,080 ratings.
Table~\ref{tab:subj_mse_8555} shows the MSE between their scores and the target VI vector.
The proposed methods generally outperformed VIC-base in the single-VI conditions, except in VIC-srf's Mod.~+3 condition on dimension I.
In the multiple-VI modulation, our methods reduced the MSE from 1.15 (VIC-base) to 1.04 (dis) and 0.92 (srf), consistent with the objective VIE-based RVI-MSE in Table~\ref{tab:exp_comparison}.
 
Second, we conducted a Mean Opinion Score (MOS) test for audio quality with 30 native or native-level English speakers (1: Bad – 5: Excellent).
The setup followed the controllability test in Table~\ref{tab:subj_mse_8555}, but including the no-modulation condition (Mod.~+0) and using 13 sentences per condition.
Each participant rated 1,014 samples, yielding 30 ratings per sample.
Table~\ref{tab:mos_speaker_8555} shows the results.
Despite GT-comparable UTMOS (Table~\ref{tab:exp_comparison}), MOS stayed below 3.8, likely due to interface bias \cite{cooper2024review}, as ``Good'' examples in our instruction anchored raters to ``Good (4)".
Audio quality showed no consistent degradation against VIC-base in either single-VI (VIC-dis: five improvements vs.\ four degradations; VIC-srf: four vs.\ four) or multiple-VI conditions. 
This comparable performance suggests our methods could improve controllability without sacrificing audio quality.

\begin{table}[t]
\centering
\footnotesize
\caption{Subjective controllability results for speaker 8555, in terms of MSE ($\pm$ 95\% CI). Bold: best, underline: second-best.}
\label{tab:subj_mse_8555}
\renewcommand{\arraystretch}{.5}
\setlength{\tabcolsep}{4pt}
\setlength{\aboverulesep}{0.3ex}
\setlength{\belowrulesep}{0.3ex}
\vspace{-3mm}
\begin{tabular}{@{}llccc@{}}
\toprule
\textbf{VI} & \textbf{Condition} & \textbf{VIC-base} & \textbf{VIC-dis} & \textbf{VIC-srf} \\
\midrule
\multirow{2}{*}{A} & Mod. -3 & $10.76 \pm 1.14$ & $\underline{8.33 \pm 1.22}$ & $\bm{7.30 \pm 1.15}$ \\
 & Mod. +3 & $3.77 \pm 0.83$ & $\bm{2.57 \pm 0.44}$ & $\underline{2.58 \pm 0.47}$ \\
\midrule
\multirow{2}{*}{B} & Mod. -3 & $0.81 \pm 0.24$ & $\underline{0.46 \pm 0.10}$ & $\bm{0.38 \pm 0.11}$ \\
 & Mod. +3 & $0.66 \pm 0.16$ & $\bm{0.40 \pm 0.10}$ & $\underline{0.46 \pm 0.10}$ \\
\midrule
\multirow{2}{*}{F} & Mod. -3 & $1.62 \pm 0.34$ & $\bm{1.37 \pm 0.31}$ & $\underline{1.48 \pm 0.62}$ \\
 & Mod. +3 & $\bm{2.15 \pm 0.44}$ & $2.22 \pm 0.45$ & $\underline{2.17 \pm 0.33}$ \\
\midrule
\multirow{2}{*}{I} & Mod. -3 & $9.42 \pm 1.19$ & $\underline{8.22 \pm 0.90}$ & $\bm{6.92 \pm 0.76}$ \\
 & Mod. +3 & $\underline{8.25 \pm 0.81}$ & $\bm{7.54 \pm 0.75}$ & $9.27 \pm 0.86$ \\
\midrule
\multicolumn{2}{@{}l}{\textbf{Multiple VIs}} & $1.15 \pm 0.10$ & $\underline{1.04 \pm 0.10}$ & $\bm{0.92 \pm 0.09}$ \\
\bottomrule
\end{tabular}
\vspace{0.1cm}
\caption{MOS for audio quality ($\pm$ 95\% CI) for speaker 8555. Bold/ $\dagger$: significantly better/worse than baseline ($p < 0.05$).}
\label{tab:mos_speaker_8555}
\vspace{-3mm}
\renewcommand{\arraystretch}{.5}
\setlength{\tabcolsep}{4pt}
\begin{tabular}{@{}llccc@{}}
\toprule
\textbf{VI} & \textbf{Condition} & \textbf{VIC-base} & \textbf{VIC-dis} & \textbf{VIC-srf} \\
\midrule

\multirow{3}{*}{A} & Mod. -3 & 3.31 $\pm$ 0.10 & 3.42 $\pm$ 0.11 & \textbf{3.75 $\pm$ 0.11} \\
& Mod. +0 & 3.31 $\pm$ 0.12 & \textbf{3.46 $\pm$ 0.11} & \textbf{3.44 $\pm$ 0.12} \\
& Mod. +3 & 3.44 $\pm$ 0.12 & 3.21 $\pm$ 0.11$^{\smash{\dagger}}$ & 2.85 $\pm$ 0.11$^{\smash{\dagger}}$ \\
\midrule
\multirow{3}{*}{B} & Mod. -3 & 3.64 $\pm$ 0.11 & 3.42 $\pm$ 0.13$^{\smash{\dagger}}$ & 3.46 $\pm$ 0.11$^{\smash{\dagger}}$ \\
& Mod. +0 & 3.37 $\pm$ 0.12 & \textbf{3.68 $\pm$ 0.11} & \textbf{3.52 $\pm$ 0.11} \\
& Mod. +3 & 3.44 $\pm$ 0.11 & 3.48 $\pm$ 0.12 & 3.34 $\pm$ 0.11 \\
\midrule
\multirow{3}{*}{F} & Mod. -3 & 3.44 $\pm$ 0.10 & 3.54 $\pm$ 0.11 & 3.43 $\pm$ 0.11 \\
& Mod. +0 & 3.34 $\pm$ 0.13 & \textbf{3.73 $\pm$ 0.10} & \textbf{3.62 $\pm$ 0.11} \\
& Mod. +3 & 3.46 $\pm$ 0.12 & 3.18 $\pm$ 0.13$^{\smash{\dagger}}$ & 3.42 $\pm$ 0.10 \\
\midrule
\multirow{3}{*}{I} & Mod. -3 & 3.51 $\pm$ 0.10 & 3.34 $\pm$ 0.11$^{\smash{\dagger}}$ & 3.29 $\pm$ 0.11$^{\smash{\dagger}}$ \\
& Mod. +0 & 3.49 $\pm$ 0.11 & \textbf{3.68 $\pm$ 0.10} & 3.26 $\pm$ 0.11$^{\smash{\dagger}}$ \\
& Mod. +3 & 3.26 $\pm$ 0.12 & \textbf{3.59 $\pm$ 0.11} & 3.18 $\pm$ 0.13 \\
\midrule
\multicolumn{2}{@{}l}{\textbf{Multiple VIs}} & 3.71 $\pm$ 0.11 & 3.28 $\pm$ 0.12 & 3.60 $\pm$ 0.12 \\
\bottomrule
\end{tabular}
\vspace{-0.65cm}
\end{table}

\vspace{-0.3cm}
\section{Conclusion}
\label{sec:conclusion}
\vspace{-0.2cm}
In this paper, we addressed two challenges in VIC: the lack of a public corpus and impression leakage. 
To address the first challenge, we introduced LibriTTS-VI, a public VI corpus.
For the second, we hypothesized that using a single reference for speaker and VI conditioning causes impression leakage.
To mitigate this, we proposed: 1) disentangled training with two utterances from the same speaker to individually provide speaker identity and the target VI, and 2) a reference-free method controlling VI without reference audio.
Our experiments showed that these methods improve controllability while maintaining high synthesis quality.
Notably, a comparison with an LLM-based TTS revealed imprecise numerical control and entanglement of text semantics and VI, which we effectively mitigated.


\section{Acknowledgments}
We thank Keiichi Osako and Fernando Villavicencio for their valuable feedback on the manuscript.

\section{Generative AI Use Disclosure}
Gemini 3 was used to support translation, paraphrasing, grammar checking, and the identification of typographical errors.
All output was reviewed and verified by the authors, who assume full responsibility for the content of the paper.

\bibliographystyle{IEEEtran}
\bibliography{mybib}

\end{document}